\documentclass[twocolumn,showkeys,preprintnumbers,longbibliography,amsmath,amssymb,epsfig,floatfix,prb]{revtex4-2}
\usepackage{array}
\usepackage[table, svgnames]{xcolor}
\usepackage{xcolor}
\usepackage{tabu}
\usepackage{amsmath}
\usepackage{mathtools}
\usepackage{physics}
\usepackage[version=4]{mhchem}
\usepackage[english]{babel}
\usepackage[utf8]{inputenc}
\DeclareMathAlphabet{\altmathcal}{OMS}{cmsy}{m}{n}
\usepackage{siunitx}
\usepackage{multirow}
\usepackage{amsfonts}
\usepackage{amssymb}
\usepackage{calligra}
\usepackage{calrsfs}
\DeclareMathAlphabet{\mathcalligra}{T1}{calligra}{l}{m}
\usepackage{epsfig}
\usepackage{graphicx}
\usepackage{dcolumn}
\usepackage{color}
\usepackage{natbib}  
\usepackage{hyperref}
\usepackage{breakurl}
\hypersetup{colorlinks=true, citecolor=blue, urlcolor=blue, linkcolor=blue}
\usepackage{bm}
\usepackage{booktabs}
\usepackage{mathtools}
\newcolumntype{C}{>{$}c<{$}}

\newcolumntype{L}[1]{>{\raggedright\arraybackslash}p{#1}}
\newcolumntype{C}[1]{>{\centering\arraybackslash}p{#1}}
\newcolumntype{R}[1]{>{\raggedleft\arraybackslash}p{#1}}

\definecolor{nblue}{RGB}{142, 68, 173}

\usepackage[table]{xcolor} 
\definecolor{LightCyan}{rgb}{0.88,1,1}
\usepackage{graphicx}      
\usepackage{dcolumn}       
\usepackage{ifthen}

\usepackage{amsmath}
\usepackage{mathtools}
\usepackage{amsfonts}
\usepackage{amssymb}
\usepackage{physics}       
\usepackage{bm}            
\usepackage{calrsfs}
\DeclareMathAlphabet{\altmathcal}{OMS}{cmsy}{m}{n}
\DeclareMathAlphabet{\mathcalligra}{T1}{calligra}{l}{m}

\usepackage[utf8]{inputenc}
\usepackage[english]{babel}
\usepackage{ulem}
\usepackage{siunitx}
\usepackage{multirow}
\usepackage{booktabs}

\begin{document}
\title{All-Electric Topological Phase Transitions in Proximity-Coupled Bilayer \ce{MnBi2Te4} Heterostructures}
\author{Basavaraja G} 
\author{Mukul Kabir}
\email{mukul.kabir@iiserpune.ac.in}
\affiliation{Department of Physics, Indian Institute of Science Education and
Research, Pune 411008, India}

\begin{abstract} 
The intrinsic magnetic topological insulator \ce{MnBi2Te4}, in the two-dimensional limit, hosts thickness dependent axion and quantum anomalous Hal (QAH) insulating states governed by antiferromagnetic interlayer coupling. However, controlled interconversion between these phases typically requires extreme external magnetic fields exceeding 9 T, limiting practical tunability. Using complementary first-principles calculations and effective Hamiltonian modeling, we demonstrate a field-free, reversible mechanism to engineer topological phase transitions by exploiting magnetic proximity at the interfaces with a ferromagnetic insulator. Gate-tunable magnetic anisotropy within the ferromagnetic insulator dynamically modulates the proximity-induced exchange bias, enabling all-electric switching of interlayer coupling and band topology in ultrathin \ce{MnBi2Te4}. Crucially, long-range Heisenberg Monte Carlo simulations reveal that the magnetic ordering temperature of the encapculated \ce{MnBi2Te4} film is dramatically elevated. By eliminating the high-field requirement and simultaneously improving thermal stability, this gate-tunable paradigm solves a critical bottleneck in topological physics and offers a viable route toward scalable, high-temperature topological electronics.
\end{abstract}
\maketitle

\section{Introduction} 
Topological phase transitions, characterized by changes in global topological indices such as the Berry phase, Chern number, and $\mathbb{Z}_2$ invariant, offer a pathway to novel quantum states beyond the Landau-Ginzburg symmetry-breaking paradigm~\citep{RevModPhys.82.3045,RevModPhys.83.1057}. These transitions, tunable via spin-orbit coupling~\citep{science.1133734,PhysRevLett.100.096407,nphys1270,science.aaa9297}, strain~\citep{s41563-021-01004-4}, doping~\citep{science.1234414,10.1038/nmat4204,10.1038/nmat4855,sciadv.aao1669}, and electromagnetic fields~\citep{s41563-019-0573-3,s41467-021-25002-x}, hold promise for applications in dissipationless spintronics and fault-tolerant quantum computing.

Ferromagnetic (FM) order in topological insulators (TI) breaks time-reversal symmetry (TRS), opening a gap in the Dirac surface states and generating dissipationless chiral edge modes. This leads to quantized Hall conductivity without Landau levels~\citep{s42254-018-0011-5,adma.202102427,science.1234414,10.1038/nmat4204}, a hallmark of the quantum anomalous Hall (QAH) effect predicted by the Haldane model~\citep{PhysRevLett.61.2015}. In contrast, antiferromagnetic (AFM) order or alternating magnetization drives the system into an axion insulator phase with fully gapped surfaces and a zero Hall plateau, corresponding to a half-quantized surface QAH response~\citep{10.1038/nmat4855,sciadv.aao1669}. Though first realized in magnetic-ion-doped TIs, these quantum states remain limited by ultralow operational temperatures of tens of milli-Kelvin and reduced quantization fidelity~\citep{science.1234414,10.1038/nmat4204,10.1038/nmat4855,sciadv.aao1669}.

Intrinsic magnetic topological insulators (MTIs) overcome these challenges by providing spontaneous magnetic order at elevated temperatures, enhanced magnetic homogeneity, and reduced disorder~\citep{Gong_2019,sciadv.aaw5685,s41586-019-1840-9}. Their van der Waals  (vdW) crystal structure and layer-dependent magnetism enable precise control of topological quantum phases, making them attractive for device applications. Odd-layer films with AFM stacking exhibit the QAH effect due to uncompensated magnetization~\citep{science.aax8156,s41563-019-0573-3}, whereas even-layer films realize axion insulator  state~\citep{PhysRevX.11.011003,s41586-021-03679-w}. However, transitions between these quantum phases typically require magnetic fields exceeding 9 Tesla~\citep{s41563-019-0573-3,s41467-021-25002-x}. Such requirements hinder scalability, underscoring the importance of developing energy-efficient control mechanisms for practical implementation. 

Here, we demonstrate electrically tunable topological phase transitions between the QAH and axion insulator states in bilayer \ce{MnBi2Te4} sandwiched between a two-dimensional (2D) ferromagnetic insulator (FMI). By modulating carrier density via electrostatic gating, the spin orientation of the proximate 2D FMI is controlled, enabling reversible switching between quantum phases. First-principles calculations~\citep{PhysRevB.47.558, PhysRevB.54.11169,PhysRevB.50.17953,PhysRevLett.77.3865,PhysRevB.57.1505,10.1063/1.3382344,PhysRevB.13.5188,PhysRevLett.115.036402,PhysRevB.65.035109,  Pizzi_2020, WU2018405}, effective Hamiltonian modeling, and Monte Carlo simulations~\citep{evans2014atomistic}, detailed in the Supporting Information, establish a low-power pathway for phase engineering.

\begin{figure*}[t]
\includegraphics[scale=0.12]{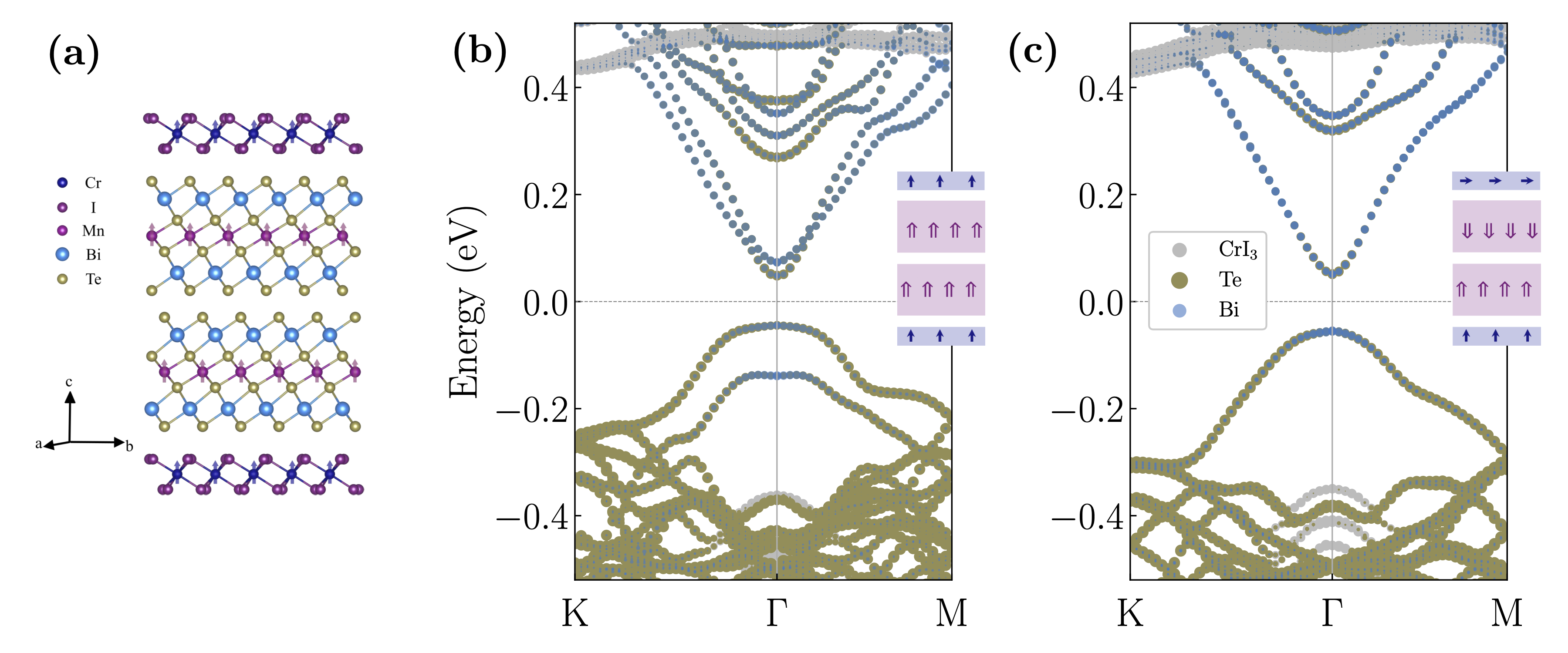}
\caption{
Controlling interlayer antiferromagnetic coupling in bilayer MTI via magnetic proximity to 2D FMI. 
(a) Schematic of a sandwich heterostructure \ce{CrI3}/2SL-\ce{MnBi2Te4}/\ce{CrI3}, where proximity-induced exchange at the FMI/MTI interfaces is used to tune topological invariants by controlling the interlayer magnetic coupling in the MTI.   
(b) Ferromagnetic exchange at the FMI/MTI interface stabilizes ferromagnetic interlayer coupling in 2SL-\ce{MnBi2Te4} [inset, ($\uparrow$/$\Uparrow \cdot \Uparrow$/$\uparrow$)], yielding the corresponding band structure.   
(c) Upon electrically tuning the magnetic anisotropy of \ce{CrI3} from out-of-plane to in-plane, the interlayer coupling in the MTI reverses to antiferromagnetic order (inset), and the corresponding band structure for the ($\uparrow$/$\Uparrow \cdot \Downarrow$/$\rightarrow$) configuration is shown.
In both configurations, the electronic structure of the MTI remains largely intact, with \ce{CrI3} bands lying well away from the Fermi level.
}
\label{fig:fig1}
\end{figure*}

\section{Results and Discussion}
\subsection{Bulk and bilayer MTI} 
Bulk \ce{MnBi2Te4} crystallizes in the rhombohedral $R\bar{3}m$ structure, composed of  {Te-Bi-Te-Mn-Te-Bi-Te} septuple layers [SLs, Figure~\ref{fig:fig1}(a)]. The optimized lattice parameters, $a = 4.37$ \AA\, $c = 40.28$ \AA\, show excellent agreement with experiments (Supporting Information)~\citep{lee2013crystal,zeugner2019chemical}. It exhibits A-type AFM order, characterized by FM \ce{Mn^{2+}} triangular layers coupled antiferromagnetically across the vdW gap mediated by long-range \ce{Mn}-$d_{z^2}$\ce{-Te}-$p_z$\ce{-Bi}-$p_z$ superexchange, while in-plane FM order arises from $90^\circ$ \ce{Mn-Te-Mn} superexchange. Easy-axis anisotropy stabilizes the magnetic order, and classical Heisenberg Monte Carlo simulations yield a N\'eel temperature $T_{\rm N}$ of 31 K (Table~\ref{exchange}), consistent with experiments~\citep{s41586-019-1840-9,zeugner2019chemical}. This picture persists in the bilayer limit, with $J_1 =$ \SI{243}{\micro\electronvolt} and $J_\perp =$ \SI{-36}{\micro\electronvolt}, although a reduced anisotropy of $0.3A_z^{\rm bulk}$ lowers $T_{\rm N}$ to 25.5 K (Table~\ref{exchange}), matching experimental observations~\citep{PhysRevX.11.011003}.

The interplay of magnetism and topology in bulk and thin \ce{MnBi2Te4} drives tunable quantum phases governed by interlayer magnetic coupling~\citep{s41586-019-1840-9,sciadv.aaw5685,science.aax8156,s41467-021-25002-x,s41563-019-0573-3,PhysRevX.11.011003,s41586-021-03679-w}. Intrinsic A-type AFM order breaks TRS $\altmathcal{T}$ while preserving the composite symmetries $\altmathcal{S} = \altmathcal{T} u_{1/2}$ and $\altmathcal{S}' = \altmathcal{P}_o\altmathcal{T}$, resulting in an AFM topological insulator with $\mathbb{Z}_2=1$, defined with respect to $\altmathcal{S}'$, on the $k_z=0$ plane~\citep{s41586-019-1840-9,PhysRevLett.122.206401}. Here $u_{1/2}$ is the half-translation between AFM aligned \ce{Mn} layers, and $\altmathcal{P}_o$ denotes the inversion center at the vdW gap.  Transitioning to FM interlayer coupling breaks both $\altmathcal{T}$ and $\altmathcal{S}'$ symmetries, driving the system into a Weyl semimetal phase~\citep{sciadv.aaw5685,s41586-019-1840-9}.

In the quasi-2D limit, \ce{MnBi2Te4} hosts layer-dependent topological phases, where surface \ce{Mn}-spin orientations dictate half-quantized Hall conductance of $\pm e^2/2h$. Odd-layer films break both $\altmathcal{T}$ and $\altmathcal{S}'$, realizing a QAH state with Chern number $\altmathcal{C}=1$, whereas even-layer films preserve $\altmathcal{S}'$ and stabilize an axion insulator with $\altmathcal{C}=0$. Achieving reversible switching between these phases within a fixed thickness device remains challenging. Bilayer (2SL) \ce{MnBi2Te4} could support both phases if the interlayer coupling $J_\perp$ were tunable between AFM and FM. However, existing approaches, using external fields~\citep{s41565-018-0121-3,s41563-021-01070-8}, gating~\citep{PhysRevLett.125.047202}, pressure~\citep{s41563-019-0506-1,s41563-019-0505-2},  or structural engineering~\citep{10.1038/s41567-019-0651,PhysRevB.99.144401,acs.nanolett.4c06419}, are typically irreversible or drive the system metallic, precluding quantized responses. To overcome this limitation and suppress electronic reconstruction, we engineer a vdW heterostructure, \ce{CrI3}/2SL-\ce{MnBi2Te4}/\ce{CrI3} [Figure~\ref{fig:fig1}(a)], where proximity to a gate-tunable 2D FMI enables dynamic and reversible control of interlayer magnetism, and electrically switchable topological phases. 

\subsection{FMI/MTI/FMI heterostructures and gate-tunable proximity coupling} 
The 2D magnetic insulator exerts a strong FM exchange bias, aligning the \ce{Mn} spins parallel to those of \ce{Cr}, without perturbing the \ce{MnBi2Te4} electronic structure (Supporting Information). The interfacial exchange coupling originating from the long-range superexchange interaction between \ce{Mn^{2+}} and \ce{Cr^{3+}} ions, mediated through \ce{Mn}-$t_{2g}^3-$\ce{Cr}-$e_g^0$ and \ce{Mn}-$e_{g}^2-$\ce{Cr}-$e_g^0$ channels, both favouring FM alignment. In the \ce{CrI3}/2SL-\ce{MnBi2Te4} heterostructure, the easy-axis \ce{CrI3} pins the adjacent \ce{Mn} layer ferromagnetically, while the interlayer coupling within the 2SL-\ce{MnBi2Te4} remains AFM, though its strength is reduced to $J_\perp = -$ \SI{22}{\micro\electronvolt} due to the proximity effect (Table~\ref{exchange}). Owing to the short-range nature of the interfacial exchange interaction and the substantial thickness of the septuple layers, the proximity effect does not propagate across the stack or modify the spin orientation of non-adjacent layers.  The interfacial coupling remains robust across different commensurate supercells, such as the ($\sqrt{3}\times\sqrt{3}$)-2SL-\ce{MnBi2Te4}/($1\times1$)-\ce{CrI3} and  ($3\times3$)-2SL-\ce{MnBi2Te4}/($2\times2$)-\ce{CrI3} configurations. In both cases, lattice-mismatch-induced strain is accommodated within the \ce{CrI3} layers without altering the qualitative physics. 

The short-range nature of the proximity effect preserves the intrinsic interlayer AFM coupling in 2SL-MBT. To overcome this limitation and enhance magnetic control, we engineer a sandwich heterostructure, \ce{CrI3}/2SL-\ce{MnBi2Te4}/\ce{CrI3}, in which interfacial FM \ce{Mn-Cr} exchange collectively drives the interlayer coupling in 2SL-MBT from AFM to FM, yielding $J_\perp = $ \SI{96}{\micro\electronvolt} [Figure~\ref{fig:fig1}(a), Table~\ref{exchange}]. The orbital-resolved band structure of the sandwich heterostructures [Figure~\ref{fig:fig1}(b) and ~\ref{fig:fig1}(c)] reveals a clean superposition of the \ce{CrI3} and \ce{MnBi2Te4} bands, confirming the absence of interfacial charge transfer. Consequently, \ce{CrI3} serves mainly as a source of proximity-induced exchange, while the electronic topology remains dictated by the MBT film. Crucially, the finite band gap and the characteristic band inversion of MBT persist, preserving the fundamental requirements for non-trivial topology.

\begin{figure}[t]
\includegraphics[scale=0.08]{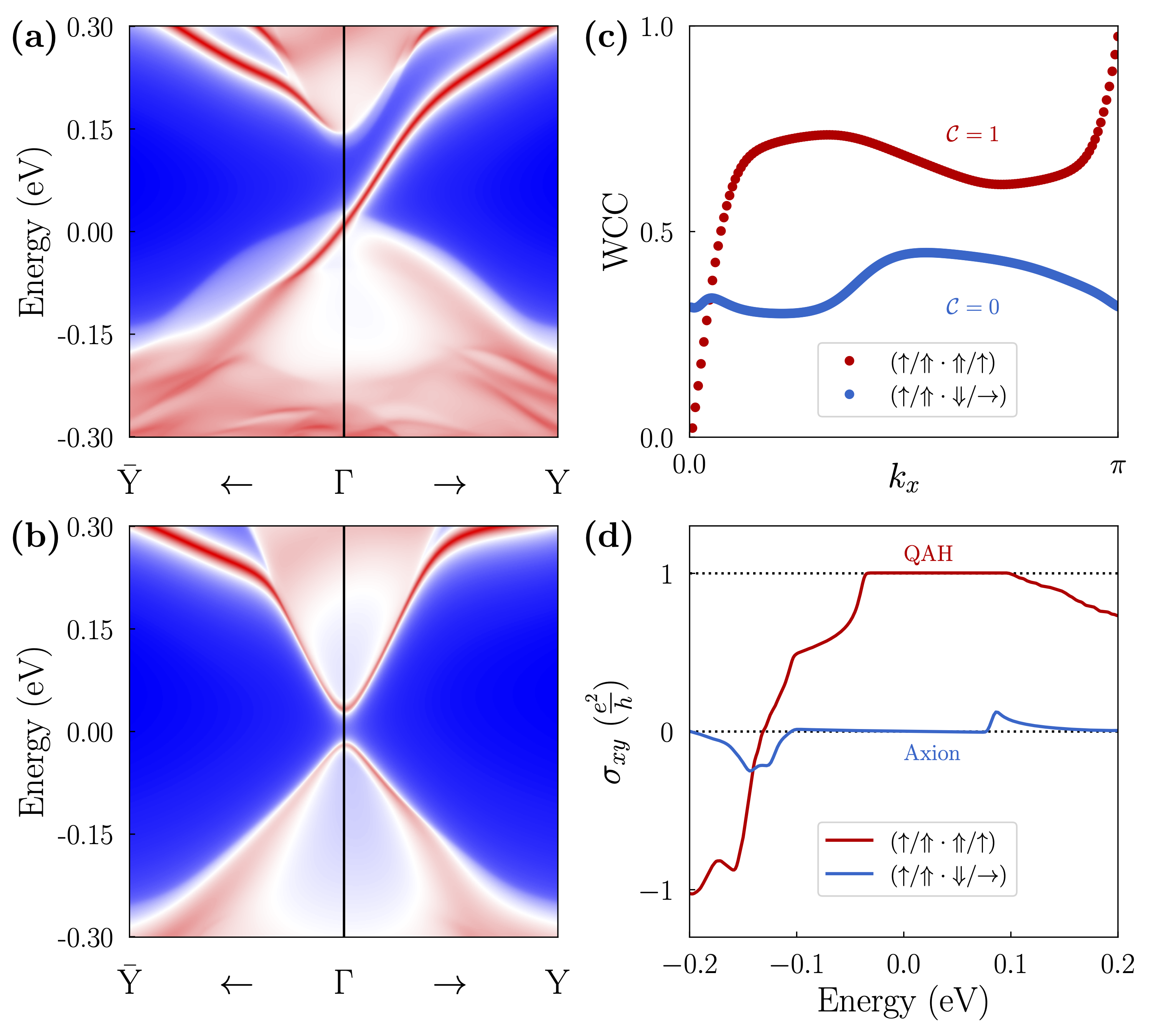}
\caption{
Switching topological invariants via gate-tunable proximity exchange. Interlayer magnetic coupling controls the surface-state topology of the MTI.
(a) The ferromagnetic ($\uparrow/\Uparrow \cdot \Uparrow/\uparrow$) configuration hosts a single chiral edge mode with $\altmathcal{C}=1$.
(b) Switching the FMI anisotropy to the in-plane direction preserves a gapped surface spectrum in the ($\uparrow/\Uparrow \cdot \Downarrow/\rightarrow$) configuration. While $\altmathcal{C}=0$ for this state, the Fu-Kane $\mathbb{Z}_2$ invariant remains nontrivial. 
(c) The calculated evolution of WCC along closed paths in the BZ confirms the corresponding topological classification. 
(d) The anomalous Hall conductivity, obtained by integrating the Berry curvature over BZ, is quantized at $\sigma_{xy}=e^2/h$ for ferromagnetic interlayer coupling, indicating a QAH state. In contrast, antiferromagnetic coupling yields a zero-Hall plateau characteristic of the axion insulator.  
}
\label{fig:fig2}
\end{figure}

At finite thickness, the breaking of composite $\altmathcal{T}u_{1/2}$ symmetry, together with out-of-plane magnetization, opens a surface Dirac gap of 98 meV, with the Fermi level positioned within the gap [Figure~\ref{fig:fig2}(a)]. The ($\uparrow$/$\Uparrow \cdot \Uparrow$/$\uparrow$) spin configuration breaks both $(\altmathcal{T})$ and $(\altmathcal{S}')$ symmetries, resulting spin-split bands.  Surface states computed using the iterative Green's functions reveal a single chiral edge mode connecting the valence and conduction bands, confirming a $\altmathcal{C}=1$ topology.  We determine the topological invariant by tracking the real-space evolution of Wannier charge centres (WCCs), obtained from the eigenvalues of the Wilson loop operator, which encodes the non-Abelian Berry phase accumulated along closed paths in the Brillouin zone (BZ). The resulting WCC flow exhibits a single winding for the FM ground state [Figure~\ref{fig:fig2}(c)], confirming a Chern number $\altmathcal{C} = 1$ that is consistent with an independent calculation of the Berry flux over the 2D BZ. The corresponding anomalous Hall conductivity, obtained by integrating the Berry curvature over the BZ, shows a quantized plateau $\sigma_{xy}=(e^2/h) \altmathcal{C}$ across the Fermi level [Figure~\ref{fig:fig2}(d)], demonstrating the emergence of a robust QAH state in bilayer MBT.

\begin{table}[t]
\centering
\caption{
Exchange parameters and single-ion magnetic anisotropies (in \si{\micro\electronvolt}) are exctracted from the first-principles calculations. The magnetic ordering temperatures  (Curie $T_{\text{C}}$ or N\'eel $T_{\text{N}}$ in K) are subsequently calculated from the Monte Carlo simulations.  The symbols $\Uparrow$ and $\uparrow$ denote out-of-plane (easy-axis) spin alignments in the $\text{MnBi}_2\text{Te}_4$ and $\text{CrI}_3$ layers, respectively, while $\rightarrow$ represents in-plane spin orientations within the $\text{CrI}_3$ layer. Bulk and bilayer benchmark results show excellent agreement with experimental reports~\cite{s41586-019-1840-9, PhysRevX.11.011003}.
}
\renewcommand{\arraystretch}{1.0} 
\setlength{\tabcolsep}{3.0pt} 
\begin{tabular}{l  r r r r r r}
\hline
\hline\\[-6pt]
                              & $J_1$ & $J_2$  & $J_3$ & $J_{\perp}$ & $A_z$ & $T_{\rm C/N}$ \\[4pt]
\hline  \\[-6pt]     
Bulk \ce{MBT}                                & 231  &  $-$18     & 4     & $-40$    & 184        & 31.0     \\[1pt]
2SL-\ce{MBT} ($\Uparrow \cdot \Downarrow$)                            & 243  & $-$13  &   4  & $-$36  &  55  & 25.5 \\[1pt] 
FMI/2SL ($\uparrow$/$\Uparrow \cdot \Downarrow$)                 & 193  &  88  &  11  & $-$22  & 354  & 37.0 \\[1pt]  
FMI/2SL/FMI ($\uparrow$/$\Uparrow \cdot \Uparrow$/$\uparrow$)        & 190  &  40  &  57  &  96  & 503  & 47.5 \\[1pt] 
FMI/2SL/FMI ($\uparrow$/$\Uparrow \cdot \Downarrow$/$\rightarrow$)   & 217 & 63 & 87 & $-29$ & 398 & 50.0 \\[1pt] 
\hline
\hline
\end{tabular}
\label{exchange}
\end{table}

Having unambiguously engineered the QAH phase in bilayer MBT sandwich heterostructure, we address the critical challenge of driving the system back into its native axion insulating state. This transition requires restoring the interlayer AFM coupling, a feat achieved by reorienting the magnetization of the proximate \ce{CrI3} layer. While an external magnetic field can induce a spin-reorientation transition once the Zeeman energy exceeds the magnetic anisotropy~\citep{pnas.1902100116}, we propose an energy-efficient alternative route based on electrostatic gating. Electron doping in \ce{CrI3} triggers a spin-reorientation transition from out-of-plane to easy-plane anisotropy at carrier densities of $n_e \sim 10^{14} \mathrm{cm^{-2}}$~\citep{PhysRevB.103.214411,PhysRevMaterials.6.084407}, a mechanism already validated in various 2D magnets~\citep{Su_2020,s41928-020-0427-7,adma.202008586,s41928-022-00882-z}.  A microscopic understanding follows from second-order perturbation theory in spin-orbit coupling, where the magnetic anisotropy energy is determined by matrix elements between occupied and unoccupied states. In \ce{CrI3}, the anisotropy originates primarily from iodine $p$-$p$ interactions. Progressive filling of the $p_z$ orbital near the Fermi level under electron doping modifies the dominant spin-orbit interaction channel, specifically the $\langle p_y \lvert L_x \rvert p_z \rangle$ matrix element, thereby driving the rotation of the easy axis into the plane.

To model spin-reorientation in gated-\ce{CrI3}, we impose an in-plane spin configuration, realizing the ($\uparrow$/2SL-MBT/$\rightarrow$) heterostructure. Remarkably, the in-plane \ce{Cr}-spins no longer exert a ferromagnetic exchange bias, allowing the system to reorganize into ($\uparrow$/$\Uparrow \cdot \Downarrow$/$\rightarrow$). In this configuration, the  bilayer \ce{MnBi2Te4} recovers its natural AFM interlayer order with $J_\perp = -$\SI{29}{\micro\electronvolt} [Figure~\ref{fig:fig1}(c), Table~\ref{exchange}].  This AFM state with out-of-plane magnetization breaks time-reversal symmetry $\altmathcal{T}$  but preserves the composite symmetries $\altmathcal{S}$ and $\altmathcal{S}'$, thereby enforcing Kramers degeneracy at all momenta. Consequently, it exhibits a gapped surface state devoid of chiral edge modes [Figure~\ref{fig:fig2}(b)], yielding $\altmathcal{C}=0$ as confirmed by the WCC flow  [Figure~\ref{fig:fig2}(c)]. Despite the vanishing Chern number, the Fu-Kane $\mathbb{Z}_2$ invariant remains nontrivial. Each surface hosts a half-quantized Hall conductivity of opposite sign, $\pm \frac{1}{2}(e^2/h)$, realizing an axion insulating state characterized by a zero-Hall conductance plateau [Figure~\ref{fig:fig2}(d)].  Thus, gate-tunable spin reorientation in \ce{CrI3} enables a reversible switch between QAH and axion insulating phases in MBT, eliminating the need for high magnetic fields~\citep{s41563-019-0573-3,s41467-021-25002-x}.  Broadly, this mechanism is not restricted to a specific FMI but generalizes to 2D ferromagnets with electrically tunable magnetic anisotropy. As demonstrated with  \ce{CrBr3} (Supporting Information), it provides a robust platform for controlling topological phases. 

\subsection{Heisenberg spin model and magnetic ordering} 
We describe the magnetism in MTI by Heisenberg spin Hamiltonian on the triangular $\mathrm{Mn^{2+}}$ ($S=5/2$) lattice with single-ion anisotropy $A_z$,  $\altmathcal{H}_{\rm spin} = -\sum_{i < j} J_{i,j} \bm{S}_i \cdot \bm{S}_j  - A_z \sum_{i} (S_i^z)^2$, where $J_{i,j}$ denotes the exchange coupling constants up to third-nearest neighbors (Table~\ref{exchange}).  While a minimal $J_1\text{--}J_\perp$ model qualitatively describes pristine MTI, capturing the phase stability of the gated heterostructure demands this long-range framework. The proximity-induced exchange field increases the out-of-plane single-ion anisotropy $A_z$ of the encapsulated bilayer by an order of magnitude while simultaneously strengthening the long-range intralayer FM constants $J_2$ and $J_3$. In the unperturbed state, the out-of-plane proximity field overwhelms the native interactions to enforce FM alignment ($J_\perp > 0$). Crucially, gating rotates the $\mathrm{CrI_3}$ spins into the plane, neutralizing this proximity field and cleanly restoring the pristine interlayer AFM coupling ($J_\perp < 0$). Consequently, Monte Carlo simulations indicate the magnetic ordering temperature of the MTI subsystem is dramatically elevated from $25.5\text{ K}$ to above $45\text{ K}$ (Table~\ref{exchange}), pinning it to the higher Curie temperature of the monolayer $\mathrm{CrI_3}$~\cite{nature22391}. This substantial enhancement in thermal stability mitigates a long-standing constraint in magnetic topological insulators, offering a viable pathway toward higher-temperature realization of the QAH and axion insulating states.

\begin{figure}[t]
\includegraphics[scale=0.11]{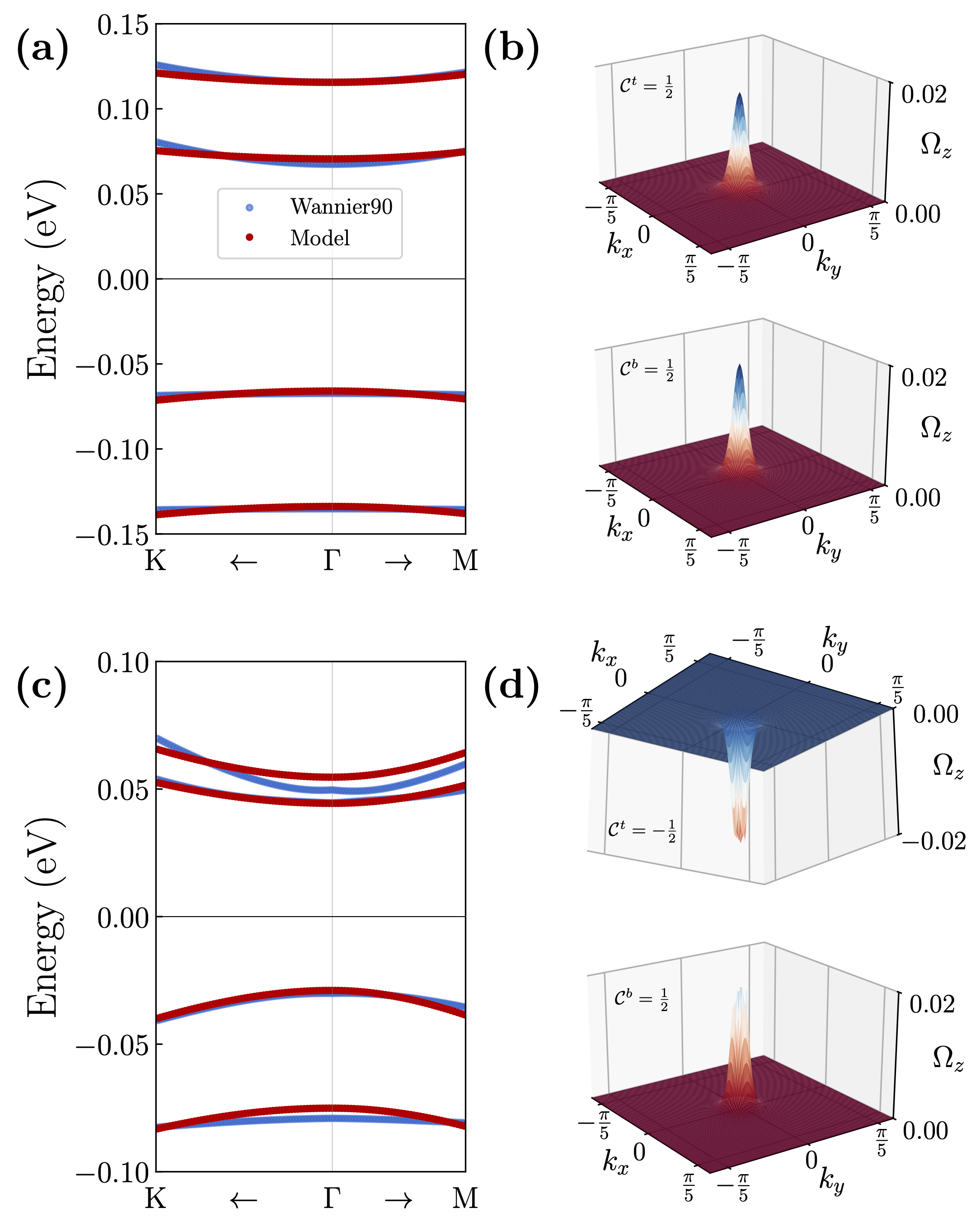}
\caption{
Low-energy $4\times4$ effective Hamiltonian at the $\Gamma$ point capturing the essential physics of the topological phase transition. 
Interpolated bands near the massive Dirac point yield the effective Hamiltonian parameters for the (a) ferromagnetic [$(M_t, M_b, V) = (102, 91, -28)$ meV] and (b) antiferromagnetic [$(M_t, M_b, V) = (-60, 42, -14)$ meV] configurations. The fitted $v_{\rm F} \sim 3.7 \times 10^5$ m/s is in good agreement with the value reported by angle-resolved photoemission spectroscopy~\citep{Gong_2019}.
(c) The layer-resolved Berry curvature for the FM configuration exhibits $\Omega_z^{t/b} > 0$, yielding half-quantized layer Hall responses on each surface with $\altmathcal{C}^{t/b} = +\tfrac{1}{2}$. Together, these contributions produce a fully quantized Hall response characteristic of the QAH state.
(d) In the AFM phase, $\Omega_z^{t}$ and $\Omega_z^{b}$ have opposite signs, yielding $\altmathcal{C}^{t/b}=\mp\tfrac{1}{2}$. The resulting cancellation produces a zero-Hall plateau, characteristic of an axion insulator. 
}
\label{fig:fig3}
\end{figure}

\subsection{Four-band model and topological phase transition} 
To corroborate the quantum phase transition obtained from first-principles calculations, we construct a low-energy effective Hamiltonian for massive Dirac surface states, $\altmathcal{H} = \altmathcal{H}_{\rm surf} + \altmathcal{H}_{\rm Z} + \altmathcal{H}_{\rm inv}$. Following the Bernevig-Hughes-Zhang formalism~\citep{science.1133734,nphys1270}, $\altmathcal{H}_{\rm surf}=v_{\rm F}(k_y\sigma_{\!x} - k_x\sigma_{\!y})$ is the 2D massless $2\times2$ Dirac Hamiltonian with $v_{\rm F}$ is Fermi velocity, $\bm{k}=(\mathit{k_x, k_y})$ measured from the Dirac point, and $\bm{\sigma} = (\sigma_{\!x}, \sigma_{\!y}, \sigma_{\!z})$ the Pauli spin matrices. The Zeeman term, $\altmathcal{H}_{\rm Z} = M\sigma_{\!z}$ captures the out-of-plane exchange field from intralayer FM ordering and proximate exchange coupling with \ce{CrI3}. Two quantum phases, realized in the  ($\uparrow$/$\Uparrow \cdot \Uparrow$/$\uparrow$) and ($\uparrow$/$\Uparrow \cdot \Downarrow$/$\rightarrow$) heterostructures, explicitly break inversion symmetry, which is represented by $\altmathcal{H}_{\rm inv} = V\sigma_{\!_0}$, with identity matrix $\sigma_{\!_0}$ and assymetric potential $V$. In the basis of top and bottom surface states $\ket{t,\Uparrow}$, $\ket{t,\Downarrow}$, $\ket{b,\Uparrow}$, and $\ket{b,\Downarrow}$, the effective Hamiltonian becomes, 
\begin{equation}
\scalebox{1.0}{$
 \altmathcal{H} = \left[
    \begin{matrix}
    h_k+M_t \sigma_{\! z}+ V\sigma_{_{\! 0}} & 0 \\
    0 & -h_k + M_b \sigma_{\! z} - V\sigma_{_{\!0}}  \\
    \end{matrix} 
    \right], 
    $}\nonumber 
\end{equation}
with  $ h_k = \left[ \begin{matrix}
                       0                         & iv_{\rm F}k_{-} \\
                      -iv_{\rm F}k_{+} & 0 \\
\end{matrix}
\right]       
$ and $k_{\pm}=k_x\pm i k_y$. Thus, $\altmathcal{H}_{\rm surf}$ describes two copies of quantum spin Hall surface states. The exchange fields $M_t$ and $M_b$ acting on the top and bottom surface states encode the interlayer magnetic configuration, distinguishing FM ($M_t, M_b > 0$) and AFM ($M_t < 0$, $ M_b > 0$) order.  The inversion-asymmetric potential is introduced as $V_{t/b} = \pm V$.

The parameters of the $4\times4$ effective Hamiltonian $\altmathcal{H}$ are determined by simultaneously fitting the four surface bands to the {\em ab initio} electronic structure [Figure~\ref{fig:fig3}(a) and ~\ref{fig:fig3}(b)].  Using these parameters, we numerically diagonalize $\altmathcal{H}$ to obtain the Bloch eigenstates $|u^\alpha_n(\mathbf{k})\rangle$, where $n$ denotes the band index and $\alpha \in \{t, b\}$ represents the layer projections.  The Berry connection for each layer-projected band is evaluated as $A^\alpha_{n,\mu}(\mathbf{k}) = i\langle u^\alpha_n(\mathbf{k}) | \partial_{k_\mu} u^\alpha_n(\mathbf{k}) \rangle,  \mu \in \{x,y\}$, from which the layer-resolved Berry curvature is computed via $\Omega^\alpha_n(\mathbf{k}) = \partial_{k_x} A^\alpha_{n,y}(\mathbf{k}) - \partial_{k_y} A^\alpha_{n,x}(\mathbf{k})$. The corresponding surface Chern number for the layer $\alpha$ is obtained by summing over the occupied bands, $\altmathcal{C}^\alpha = \sum_{n \in \text{occ}} \frac{1}{2\pi} \int_{\mathrm{BZ}} d^2k \, \Omega^\alpha_n(\mathbf{k})$. For the ($\uparrow$/$\Uparrow \cdot \Uparrow$/$\uparrow$) configuration, both surfaces exhibit positive Berry curvature $\Omega^{t/b} >0$ [Figure~\ref{fig:fig3}(c)]. This yields half-integer Chern numbers $\altmathcal{C}^{t/b}\!=\!1/2$, and corresponding half-quantized surface Hall conductances, $\sigma_{xy}^{t/b} = e^2/2h$. Their sum produces a fully quantized Hall response, $\sigma_{xy}=e^2/h (\altmathcal{C}=1)$, characterizing the QAH state. In contrast, the gated ($\uparrow$/$\Uparrow \cdot \Downarrow$/$\rightarrow$) configuration generates Berry curvature of opposite sign on the two surfaces [Figure~\ref{fig:fig3}(d)], yielding $\altmathcal{C}^{t/b}\!=\!\mp1/2$ and $\sigma_{xy}^{t/b} = \mp e^2/2h$. The resulting cancellation leads to a zero Hall response, $\sigma_{xy}=0$, realizing the axion insulating phase. This remarkable agreement between the effective model and first-principles calculations confirms that the field-free topological interconversion is driven fundamentally by the localized modulation of surface exchange fields. Furthermore, this analysis demonstrate how individual surface-projected topological invarients and half-quantized Hall response dictate the global quantum phases.

\section{Conclusion} 
We have demonstrated a field-free topological quantum phase transition between quantum anomalous Hall and axion insulator states by exploiting magnetic proximity effects at FMI/MTI/FMI interfaces. Electrically tuning the magnetic anisotropy within the FMI layers dynamically modulates the proximity-induced exchange bias, enabling reversible, all-electric control over both the interlayer coupling and the global band topology of the encapsulated MTI. This eliminates the high magnetic field requirement,  typically exceeding 9 T,  necessary for phase interconversion, which has long remained a critical bottleneck in topological physics. Ultimately, this proximity-induced switching mechanism offers a robust, low-energy paradigm for manipulating topological phases in van der Waals heterostructures and establishes an experimentally accessible platform for the deterministic control of topological invariants in scalable electronic devices.

\section{Acknowledgements}
 B. G. acknowledges support from the University Grants Commission. M. K. acknowledges support from the National Mission on Interdisciplinary Cyber-Physical Systems, Department of Science and Technology, Government of India, through the I-HUB Quantum Technology Foundation, Pune. Computational resources were provided by the PARAM Brahma Facility at IISER Pune under the National Supercomputing Mission of the Government of India.
 
 \medskip
 \textbf{Supporting Information} \par 
Supporting Information is available from the Wiley Online Library.


%

\clearpage
\onecolumngrid

\newcounter{pdfpagecount}
\setcounter{pdfpagecount}{1}

\makeatletter

\whiledo{\value{pdfpagecount} < 16}{
    \thispagestyle{empty} 
    \noindent
    \centerline{
        \includegraphics[page=\value{pdfpagecount}, width=1.00\paperwidth, height=0.95\paperheight, keepaspectratio]{Supporting_Information.pdf}
    }
    \clearpage
    
    \ifnum\value{pdfpagecount}=15
        \let\ps@plain\ps@empty
        \let\ps@headings\ps@empty
        \thispagestyle{empty}
    \fi
    
    \stepcounter{pdfpagecount}
}

\makeatother

\pagestyle{empty}

\end{document}